# A Study on the Wetting Properties of Broccoli Leaf Surfaces and their Time Dependent Self-Healing After Mechanical Damage

*Benjamin B. Rich[a], Boaz Pokroy[a,b]\**

Plants are protected from the elements by a complex hierarchical epicuticular wax layer which has inspired the creation of super-hydrophobic and self-cleaning surfaces. Although many studies have been conducted on different plant wax systems to determine the mechanisms of water repulsion hardly any have studied the recovery of the epicuticular wax layer. In the current study the wetting properties and crystallographic nature of the wax surface of Brassica oleracea var. Italica (broccoli) has been studied, as well as the time-dependent recovery of the surface after mechanical damage. It was found that the surface of the broccoli leaves is not only super-repulsive and self-cleaning in regards to water but also in regards to glycerol and formamide both of which have considerably lower surface tension values. Furthermore, it was shown that the surface properties do indeed recover after damage and that this recovery is multi-steped and strongly dependent on the recovery of the roughness of the surface.

## Introduction

To protect themselves from their environment, plants have evolved complex hierarchical surfaces called cuticles, which cover the majority of the plant's organs (notable exceptions are roots and bark).[1-3] The cuticle has many functions, of which the three most important are to serve as a transport barrier, to reduce water loss, and to reduce or increase surface wettability.[4-8] The cuticle can also provide the plant with anti-adhesive and self-cleaning properties [1,9,10], as well as protection from harmful radiation [11,12] and reduction of surface temperature.[13,14] The plant cuticle and its multitude of combined functions has acted as a major source of inspiration for scientific research and imitation attempts. It has led to the creation of bio-inspired super-hydrophobic self-cleaning surfaces derived from the Lotus phenomenon [7,15,16], as well as de-icing [17,18], and anti-biofouling surfaces.[19,20]

The cuticle is composed of a framework of insoluble high-molecular-weight polyester interspersed with soluble long-chain cuticular waxes. Coating the cuticle are epicuticular waxes, which can create 2-dimensional wax layers and 3-dimensional wax structures.[21-23] Both the wax films and the epicuticular crystals that cover the leaf surface are formed via the self-assembly of paraffin molecules.[24,25] The great variation in chemical composition of the cuticles and waxes of different plant species gives rise to a large variety of surface structure shapes.[4,26] The macrostructure of the leaf surface is also extremely diverse in the plant world, as reflected, for example, in different cell formations and surface folds, and in papillary and hair aspect ratios.[27,28]

Folding of the cuticle and the formation of 3-dimensional wax crystals on its outer surface play a critical part in the surface wettability. The underlying hydrophobic wetting properties of the wax are enhanced by the addition of a high degree of hierarchical roughness, creating a super-hydrophobic surface.[29-32]

Wetting refers to the behavior of a liquid when deposited on a solid substrate. The contact angle (CA) of the liquid on a surface depends on the surface free energy of their respective phases.[33,34] A 'super-hydrophobic' surface is defined as a surface on which the static CA of the water droplet is greater than 150°, and if the surface also has a contact angle hysteresis (CAH) below 10° it is considered to be both super-hydrophobic and self-cleaning.[35-37] The behavior of a liquid on a rough surface has been intensively studied and was modeled by Wenzel[38] and Cassie and Bexter.[39] Wenzel assumes that the surface chemistry is homogeneous; hence the effect of added roughness is described by the Wenzel equation as:

$$\cos \theta_W = r \cos \theta_Y \qquad (1)$$

Where $\theta_Y$ is the ideal Young-Laplace CA and $r$ is the roughness ratio, defined as the ratio of the true contact area to its projected area. Cassie and Bexter , in contrast, describe a rough surface as a surface having an energetic preference for the creation of air pockets, leading to the creation of a chemically heterogeneous surface. In this state the wetting is described by the Cassie-Baxter equation:

$$\cos \theta_{CB} = f_{SL}(r_f \cos \theta_Y + 1) - 1 \qquad (2)$$

Where $f_{SL}$ is the fraction of the projected area of the solid surface that is wet and $r_f$ is the roughness ration of the wet area. It is this latter model that more accurately describes the super-hydrophobic properties of some plant leaves.

In this study we focus predominantly on the wetting properties of broccoli (*Brassica oleracea* var. *italica*). This species has one of the more complex wax crystal structures and the *italica* variety has not been studied as thoroughly as some of its other family members. Broccoli is one of the cultivated cole crops, and is distinct from its close relative, the cauliflower, by their differing ontogeny at the mature harvestable stage, which determines their taxonomic classifications.[40]

Despite the great potential application of super-hydrophobic surfaces in industry, [16, 41,42] the use of self-cleaning surfaces is currently limited by the fact that the micro- and nano-scale structures on these surfaces are susceptible to wear and must therefore be constantly maintained.[43-46] Furthermore, moving water is needed in order to remove adhering particles. We hypothesize that by combining the self-healing properties and the self-cleaning via dew and fog that appear to characterize the broccoli leaf surface [27] it might be possible to circumvent these problems.

Published research on the recovery of cuticular wax has not concentrated specifically on the epicuticular wax structures or on the effects of damage and recovery on the wettability of the plant surface.[24,47] We show here that there is recovery not only of the super-hydrophobic properties of the broccoli leaf but also of its self-cleaning properties, illustrating the important biomimetic potential this specific plant-surface system.

## Experimental

Broccoli was grown from seed in soil watered with a standard Hoagland solution at pH 6.5 pH.[48] Prior to planting, the seeds were sanitized in a 1:3 bleach and water solution for 3 min and the soil was sanitized using a Tuttnauer 2540M autoclave for 1 h at 100°C. The broccoli sprouted and grew to maturity under controlled conditions at 16°C and an 8/16 hour light/dark cycle using an MRC 550 growth incubator (MRC Laboratory Products, Holon, Israel).

Static wetting properties (CA) and dynamic surface-wetting properties (CAH) were measured with an Attension Theta Tensiometer. For measurements, we used 7 µl drops of the following liquids: high purity water (Milli-Q), diiodomethane (99% pure, Sigma-Aldrich), formamide (≥99.2% EMSURE, Merck), dimethylsulfoxide (≥99.9%, Carlo Erba), ethylene glycol (EG) (≥99.5% EMSURE), triethleneglycol anhydrus pract. (Fluka Chemie), glycerol (≥99.5%, Bio-Lab), triethylene glycol monomethyl ether (95%, Sigma-Aldrich), ethanol analytical reagent grade (Bio-Lab), isopropanol (Gadot ≥99.8%), and n-hexadecane (99%, Alfa Aesar 99%). CAH was measured for water drops of 7 ± 5 µl. When measuring the CAH of the more viscous liquids (glycerol and EG), it was necessary to use a needle with a larger opening diameter to obtain larger initial liquid drop sizes. For glycerol a 3.096-mm diameter needle was used and the drops measured were 10 ± 7 µl. For EG a 2.15-mm diameter needle was used and the drops measured were 25 ± 18 µl.

Recovery processes of wetting properties and roughness were assessed on leaves harvested from mature plants. A cotton swab was used to mechanically damage the broccoli leaf surface on one side while it was still attached to the 'mother plant', and the other side was left undamaged, enabling us to obtain both a damaged and a control section **(Figure 1)**. Every 24 h specimens were cut from the mother plant, transferred to glass slides, and fixed using two-sided tape. On each leaf CA and CAH measurements were taken more than 5 times and repeated on at least 5 leaves from different plants. Using a Leica DCM 3D confocal microscope, we also took roughness measurements at least 5 times for each leaf and repeated them on at least 5 different leaves. In vivo, roughness was measured on living plants placed under the confocal microscope. Prior to mechanical damage, control values were acquired. The specimen was then damaged at the same location and time-resolved measurements were obtained at shorter intervals.

Surface imaging was performed by HR-SEM (Zeiss Ultra Plus) by employing a liquid substitution method as suggested by Ensikat and Barthlott (1993)[49] and then coating the sample with a layer of carbon approximately 5 nm thick.

For X-ray measurements the leaf samples were placed in transmission mode using a Rigaku D/teX Ultra detector. The lower surfaces of the samples were lightly washed with chloroform to eliminate the bottom surface wax. To examine the broccoli via transmission XRD with and without the epicuticular wax layer, we used glue for stripping the surface to remove the wax.[24] For this method, we used a two-component epoxy glue (3M™ Quick set epoxy adhesive) to strip the wax layer from the plant epidermis cells. The two components were mixed, the glue was spread over the leaf and left to cure it, and the extent of wax removal was determined by the time allowed for curing. After the desired curing time of either 1 or 2 h, until the glue had solidified enough to enable its removed without leaving a residue, or after approximately 12 h (for full curing), the glue was stripped off and the sample was analyzed. Using a light chloroform wash,[50,51] powder diffraction samples of cuticular wax were taken from the wax removed from numerous broccoli leaves, and were measured on ID22 at the ESRF in Grenoble, France. Approximately 30 leaf surfaces were washed, 4 times for each leaf. The wax solvent solution was then left to evaporate under a chemical hood until the wax was crystallized and the solvent was completely evaporated. In addition, sub-micron scanned synchrotron diffractions of undamaged broccoli leaf surfaces were measured on ID13 at the ESRF.

## RESULTS AND DISCUSSION
### Surface morphology

To properly evaluate the surface morphology of broccoli leaf specimens it is important that both the cell structure and the epicuticular wax structures remain undamaged throughout the preparation process as well as during imaging. The liquid substitution method[48] preserves both the cell structures and the wax epicuticular structures, and the similarity of our results using this method to those obtained using cryo-scanning electron microscopy (SEM) and optical microscopy **(Figure 2a–f)** indicate that liquid substitution is indeed a fitting procedure for the preparation of the specimens used in our research. With this method, moreover, the surface is not in contact with any medium other than air, ensuring that it always remains as close as possible to its natural state.

High-resolution (HR)-SEM imaging (Figure 2c-e) of the macro- and micro-hierarchical structures on the broccoli leaf surface shows that the upper surface consists of polygonal cells on the leaf lamina (the areas between the plant veins, Figure 2c,f) and elongated tetragonal cells on the veins (Figure 2e). Differences in cell shape between the different parts of the leaf reflect their different functions. Since veins direct the flow of water from the leaf surface their cells are specifically oriented parallel to the vein's direction, causing water droplets to take the easy route off the leaf. On the lamina, which does not possess this particular function, orientation of the cells is random. Covering the leaf's entire surface are epicuticular wax structures with an intricate dendritic rodlet shape (Figure 2d), creating a dense dendritic 'forest'. Whereas the length of these dendrites varies considerably, their diameter is approximately 112 ± 18 nm with a typical spacing of 146 ± 74 nm between them. Oblique images show that the wax structures are approximately 2 μm in height.

**Crystallographic structure**

To elucidate the crystallographic structure of the wax, we performed *synchrotron X-ray powder diffraction* on solution-washed broccoli wax powder and transmission X-ray diffraction (XRD) on the leaves. The diffractions from the broccoli wax powder, the transmission diffraction from a pristine undamaged leaf, and—as a control—the diffraction of pure nonacosane ($C_{29}H_{60}$) are presented in **Figure 3a**. Clear peaks were obtained both from the powder and from transmission diffractions that correspond to the (110) and (200) planes of the nonacosane wax powder (the major component of the broccoli wax).[4,52] These findings gave some indication of the crystalline nature of the broccoli wax, both in its natural state on the leaf and in the powdered form. The location of peaks, as well as their correlation with peaks from the pure nonacosane diffractions, indicated that the methylene-containing sub-cells of the broccoli wax have an orthorhombic lattice structure similar to that of the pure paraffin.[53-55] On the other hand, the deviations observed between the diffractions of the natural wax (from the pristine unwashed leaf) and solution-washed broccoli wax from the diffraction of the pure nonacosane indicated some deviation of the crystalline structure from that of pure nonacosane. Most prominent is the absence of the 'long spacing' peaks in the broccoli wax powder diffraction. This points to a loss of the long-range order along the *c* axis and the formation of a 'nemato-crystalline' state, which consists of a 2-dimensional order in the *ab* plane and a disordered c axis. This most probably occurs as a result of the broad distribution of alkane chain lengths and the presence of substituted alkanes in the natural wax.[25]

Another deviation from the diffraction of pure nanocosane is a shift in the peak position of the broccoli wax to a smaller 2θ, indicating an increase in the lattice parameters. The lattice parameters of pure nonacosane are *a*=4.950Å, *b*=7.44Å, and *c*=77.52Å [55] while we found (using full profile fitting) that the lattice parameters of broccoli wax are *a*=4.99Å, *b*=7.50Å, and *c*=78.78Å. The observed increase in the lattice parameters is attributable both to the longer n-alkane chains present and to side substituents (ketones and secondary alcohols) that cause the otherwise closely packed alkane chains to separate slightly.[25]

Three types of specimens were subjected to XRD: leaves with a pristine wax layer, leaves from which the wax was fully stripped by the use of cured epoxy glue, and leaves on which the wax was partially stripped using partially cured glue (Figure 3b). At low angles we found a broad peak, presumably reflecting the organic matter in the leaves, as it was observed on diffraction in all three specimen types. Comparison of the diffractions in the three types of specimens revealed that the observed sharp diffraction peaks are derived from the epicuticular wax structures rather than from some other source inside the leaves, as these peaks appeared only in the pristine leaves and not in the stripped leaves. At higher angles the observed peak was very broad in both the untouched and the partially stripped specimens, but not in the specimen that was fully stripped. This last diffrcation peak could be attributed to either an amorphous or a nanocrystalline phase located between the epicuticular wax structures and the organic leaf cells, suggesting that it probably belongs, respectively, either to the two-dimensional epicuticular wax layer or to the paraffin molecules within the cuticle itself.

To further substantiate the crystallographic nature of the wax epicuticular structures, we scanned a pristine leaf using the sub-micron synchrotron XRD on line ID13 at the ESRF. The results can be seen in Figure 3c. Given the distinct diffraction patterns that appeared on this scan, we can confirm that the broccoli wax is indeed crystalline in nature. Regrettably, owing to the orientation of the specimen it was not possible to extract any conclusive information about the relative orientations of all the wax crystals.

Despite the similarity of the crystalline structures in both the naturally occurring undamaged broccoli wax crystals and the solution-deposited broccoli wax crystals, the crystal shapes are not necessarily similar, as can be seen in **Figure 4**. Indeed, the shape of the broccoli wax that is crystallized from the solution bears closer similarity to that of the pure nonacosane than to the naturally occurring crystals, and possesses the rough lozenge-type shape characteristic of n-alkanes crystallized from solution.[52,56] This greatly strengthens the claim that while the wax chemistry does indeed have an effect on development of the natural crystal shape, the crystallizing mechanism is also extremely important in this regard, and the biological process is necessary in order to achieve the complex crystal shape seen on the surface of the broccoli leaf.

**Wetting**

To determine the wetting behavior of damaged broccoli leaves, we need to understand the initial surface properties of the undamaged wax layer. The observed behavior of water droplets on the undamaged broccoli leaf strongly suggests that the leaf surface is indeed highly hydrophobic and self-cleaning (**Figure 5** and **Figure S1**). To quantitatively analyze the leaf's wetting behavior we measured the contact angle (CA) of water and of other common solvents. The surface tension, its components, and the measured CA values of these liquids are recorded in Table 1. Furthermore the measured CA in relation to the total surface tension and the different surface tension components are plotted in **Figure 6a-c**. The average measured CA of water is 161.1 ± 0.5°, which confirms that the undamaged broccoli surface is super-hydrophobic. Figure 6a shows that the general trend is a decrease in CA with decreasing total surface tension of the liquid. There are some exceptions to this trend, however, as in the case of triethylene glycol where the CA is greater than expected, and diiodomethane, where the CA is smaller than expected.

To gain a better understanding of the surface properties, we plotted the surface tension components of liquids that show phobic tendencies towards the surface (CA >90°) against their measured CA values. The surface tension components are known for some of these liquids, but regretably not in all of them. No connection was found between a liquid's measured CA and its dispersive component (Figure 6b). On the other hand, a connection between the liquid's CA and its polar component was detectable, as the CA showed a distinct decrease with decreasing values of the latter surface tension component (Figure 6c). This data indicated that the broccoli leaf surface is strongly dispersive, in agreement with the known composition of natural broccoli wax, as pure alkanes (its major component) are apolar in nature.[59,63,64] Interestingly, although a comparison of formamide and ethylene glycol shows that their polar component values are the same, the CA of formamide is slightly higher—a finding that might be attributable to the significantly higher dispersive component of formamide. We can therefore deduce that although the surface of the broccoli leaf has a strong dispersive tendency, it nevertheless still possesses some polar properties owing to the various polar substituents (ketones and secondary alcohols) in the natural wax. This phenomenon shows that not all hydrocarbon tails of the substituted alkanes face outwards, as is typical for pure polar-substituted alkane systems,[25, 65] and that some of the polar substituents are found on the surface.

Also detected when studying the polar component in relation to the CA was the appearance of two different wetting regimes for the liquids, as denoted by their different slopes (Figure 6c). To verify the apparent existence of two different wetting regimes, we also measured the contact angle hysteresis (CAH). **Figure 7** shows the values of measured CAH in relation to the polar components of the liquids used. If we define self-cleaning surfaces as those in which the CAH is lower than 10°, we can separate the liquids into two classes: those that exhibit self-cleaning properties on the broccoli leaf and those that do not. These two classes were found to coincide, respectively, with the liquids composing the two wetting regimes shown to differ on the basis of their slopes. We deduce from the CAH results that the two wetting regimes respectively represent the Wenzel and the Cassie-Baxter wetting states on the broccoli leaf surface, where water, glycerol, and formamide are in a Cassie-Baxter state whereas ethylene glycol, DMSO, and diiodomethane are in a Wenzel state. The average CAH of water measured here was 3.6 ± 0.6°, further corroborating the assessment that the broccoli leaf surface is self-cleaning as well as super-hydrophobic.

Table 1 - Values of surface tension components of different solvents at 20°C and their contact angles on the broccoli leaf surface.

| Liquid | C.A. [deg] | $\gamma^{tot}$ [mN/m] | $\gamma^{LW}$ | $\gamma^{AB}$ | $\gamma^+$ | $\gamma^-$ |
|---|---|---|---|---|---|---|
| Water[a] | 161.1±0.5 | 72.8 | 21.8 | 51.0 | 25.5 | 25.5 |
| Glycerol[a] | 156.6±4.4 | 64.0 | 34.0 | 30.0 | 3.92 | 57.4 |
| Formamide[a] | 146.7±2.7 | 58.0 | 39.0 | 19.0 | 2.28 | 39.6 |
| Diiodomethane[a,b] | 100.5±5.6 | 50.8 | 50.8 | 0 | 0.72 | 0 |
| Ethylene Glycol[a] | 142.5±3.8 | 48.0 | 29.0 | 19.0 | 1.92 | 47.0 |
| DMSO[c] | 115.1±2.9 | 44 | 36 | 8.0 | 0.5 | 32.0 |
| Triethylene glycol[d] | 149.3±3.8 | 44.71 | n.a. | n.a. | n.a. | n.a. |
| Methoxytriglycol[e] | 80.7±4.5 | 36.4 | n.a. | n.a. | n.a. | n.a. |
| Ethanol[c] | (spreads) | 24.67 | 22.4 | 2.27 | 0.019 | 68 |
| Isopropanol[f] | (spreads) | 23.87 | 20.86 | 3.01 | n.a. | n.a. |
| n-Hexadecane[c] | (spreads) | 27.47 | 27.47 | 0 | 0 | 0 |

[a] – [57], [b] – [58], [c] – [59], [d] – [60], [e] –[61], [f] –[62]

**Roughness**

The hierarchical structure of the leaf surface (**Figure 8**) hinders the measurement of the surface roughness, as it causes the macro-roughness (due to the plant's cells) to obscure the micro-roughness (due to the epicuticular wax structures), thereby preventing us from obtaining realistic results. To overcome this problem we filtered between the surface macro-roughness of the plant cells and the micro-roughness of the wax crystals (**Figure 9**), using a robust Gaussian filter with a cut-off wavelength of 8 μm.

The roughness measured on the pristine broccoli leaf surface is presented in Table 2. To correctly measure the macro-roughness we obtained topographical images at 50× magnification so that they encompassed multiple plant cells, and images for measurements of the micro-roughness at 150× magnification.

|  | Macro-roughness [$\mu m$] | Micro-roughness [$\mu m$] |
|---|---|---|
| **Magnification** | 50X | 150X |
| **Roughness** | 3.03 ± 0.06 | 0.53 ± 0.02 |

**Table 2** - Roughness results for undamaged broccoli leaf surface

**Recovery**

To evaluate the recovery process of the broccoli leaf surface, we damaged the surface and then followed the evolution of the CA and the CAH. **Figure 10** and **Figure 11a** show that damage resulted in an immediate drop of approximately 35° in the CA, after which there was a gradual return to control values, with full recovery of the surface super-hydrophobicity after approximately 70 hours. The damage also resulted in an increase in CAH values, as seen in Figure 11b, which corresponded to a temporary elimination of the self-cleaning property. Following damage, the CAH increased by approximately 14° and then returned to control values (indicating full recovery of the self-cleaning property) after approximately 150 hours.

To understand why full recovery of the CAH took twice as long as recovery of the CA we measured the time-resolved roughness of damaged leaf surfaces. This was done by ex-vivo experiments similar to those carried out for the CA and CAH, as well as by in-vivo experiments in which the roughness was measured at 150× so that the wax crystal structures would be more prominent than the plant cells. Damage to the surface resulted in a sharp drop of approximately 0.25 μm in the micro-roughness (a drop of about 50% of the initial roughness as seen in Table 2), followed by a gradual recovery and then reaching control values after about 130 h (**Figure 12**). This data is in good agreement with the observed recovery of the CAH (Figure 11b), and indicates that the CAH is more dependent than the CA on the roughness, as the CA was restored before the roughness was fully recovered (Figure 11a). It is important to note that because we are dealing with biological systems, our samples inherently have relatively large deviations from leaf to leaf. It is also important to note that the CA, CAH, and roughness measurements were taken at different times and on different samples, and for this reason the population size was relatively large. Topographical images were taken at a magnification of 50× to better evaluate the recovery of plant cells. In this case it is interesting to compare the recovery of the macro-roughness (via waviness filter) to that of the micro-roughness (via roughness filter) (**Figure 13a** and b, respectively).

The in-vivo experiments enabled us to conduct measurements at smaller time scales to better determine the nature of the recovery of roughness. The trend at the beginning of the recovery process differs from the trend observed at longer time intervals. Recovery of the micro-roughness is relatively stable, and starts only after about 15 hours **(**Figure 13b**)**. The macro-roughness, on the other hand, shows a dramatic increase in recovery immediately after the damage is inflicted, which then stabilizes for about 20 hours before resuming its increase (Figure 13a). We attributed this sharp rise in initial roughness to re-expansion of the flattened plant cells. Since it is known that a hierarchical surface texture is needed in order to support super-hydrophobic properties,[29-32] recovery of both the macro-roughness and the micro-roughness is critically important. Also notable in connection to this is the finding that the micro-roughness begins to recover only after the initial recovery of the macro-roughness (i.e., of the plant cells recover). Given these observation, we suggest that if the damage, whether mechanical or otherwise, is severe enough to harm the cell structure, we may not see any recovery at all. To corroborate the roughness data we obtained time-dependent high-resolution scanning electron microscopy (HR-SEM) images of the broccoli surfaces. These images revealed the damage-induced destruction of the wax crystal structures, followed by their growth until a dense layer of wax structures covered the entire leaf surface (**Figure 14**). The recovering crystal structures started as platelets, becoming increasingly complex until finally they began to form dendritic rodlets similar to those seen before any damage had been incurred (**Figure 15**). The observed specific orientation of crystal growth may be attributable to the ongoing growth of the leaf which, as discussed by Neinhuis, Koch and Barthlott,[47] may cause anisotropic tension in the leaf that induces the preferred orientation. It would be interesting to observe if the wax plate orientation is random in regards to the leaf or is perpendicular or parallel to the length axis of the leaf. From the HR-SEM images it is clear that to achieve a self-cleaning surface it is necessary to have a very dense system of asperities while to create a surface with low wettability the density is a less important factor.

## SUMMARY AND CONCLUSIONS

In this work we characterized the broccoli leaf surface with respect to its morphology, roughness, wetting behavior, and the recovery of these properties after mechanical damage. We also studied the crystalline structure of the broccoli wax. Using a liquid substitution process and HR-SEM, we found that the broccoli leaf surface is composed of polygonal cells covered by a dense 'forest' of dendritic rodlet wax crystals. Together these entities create a hierarchical structure characterized by both macro- and micro-roughness. We found that this surface structure enables the broccoli leaves not only to demonstrate both super-hydrophobic and self-cleaning properties with water but also to exhibit remarkably high contact angles and contact angle hysteresis with liquids of lower surface tension such as formamide. Analysis of the collected CA data revealed both a dispersive and a polar component, indicating the presence of polar groups intermixed with n-alkane hydrocarbon tails on the wax surface.

Crystallographic characterization showed conclusively that that the broccoli wax is crystalline in nature, possessing orthorhombic sub-cells that form a nemato-crystalline superstructure owing to its diverse and complex chemical composition. This study also highlighted the importance of the biological crystallization process in the creation of these complex wax crystal structures. In addition, it revealed the existence of a second layer, composed of nano-crystalline or amorphous wax, located between the epicuticular wax layer and the plant cells themselves.

Upon examining the recovery of the leaf surface after mechanical damage we found that damage to the broccoli leaf causes a deterioration in its wetting properties, owing mainly to destruction of the wax structures and hence a reduction in the micro-roughness. Following damage, both the static contact angle (CA) and the dynamic contact angle (CAH) recovered fully, albeit at different rates. Recovery of the plant cells was followed by recovery of the wax crystal structures, and hence of the hierarchical surface roughness.

We conclude that the hierarchical structure of the broccoli leaf, combining both macro- and micro-roughness, is necessary for its super-hydrophobic and self-cleaning properties. Furthermore the broccoli leaf surface is capable of self-healing after being damaged. We can also state with conviction that the CAH is more dependent than the CA on surface roughness and a high density of asperities. This means that in designing a surface intended to be super-hydrophobic but not self-cleaning, a smaller degree of roughness is needed and therefore a simpler fabrication process can be employed.

## Conflicts of interest

The authors declare that there are no conflicts of interests.

## Acknowledgements

The research leading to these results has received funding from the European Research Council under the European Union's Seventh Framework Program (FP/2013-2018)/ERC Grant Agreement No. 336077. The authors thank Dr. Davide Levy for help with the crystallographic data. The authors are also indebted to the ESRF ID22, for use and support of the high-resolution powder beamline as well as ID13 for use and support of the sub-micron scanning beamline.

## AUTHOR INFORMATION

**Corresponding Author**

* Email - bpokroy@tx.technion.ac.il

**Images** (first mention it text is in **bold**)

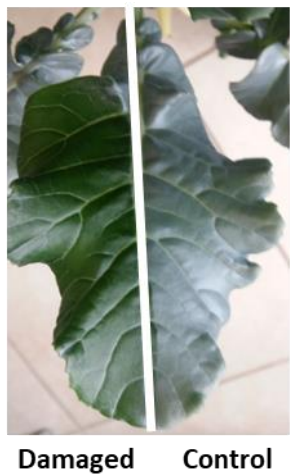

**Figure 1** - Example of broccoli leaf sample.

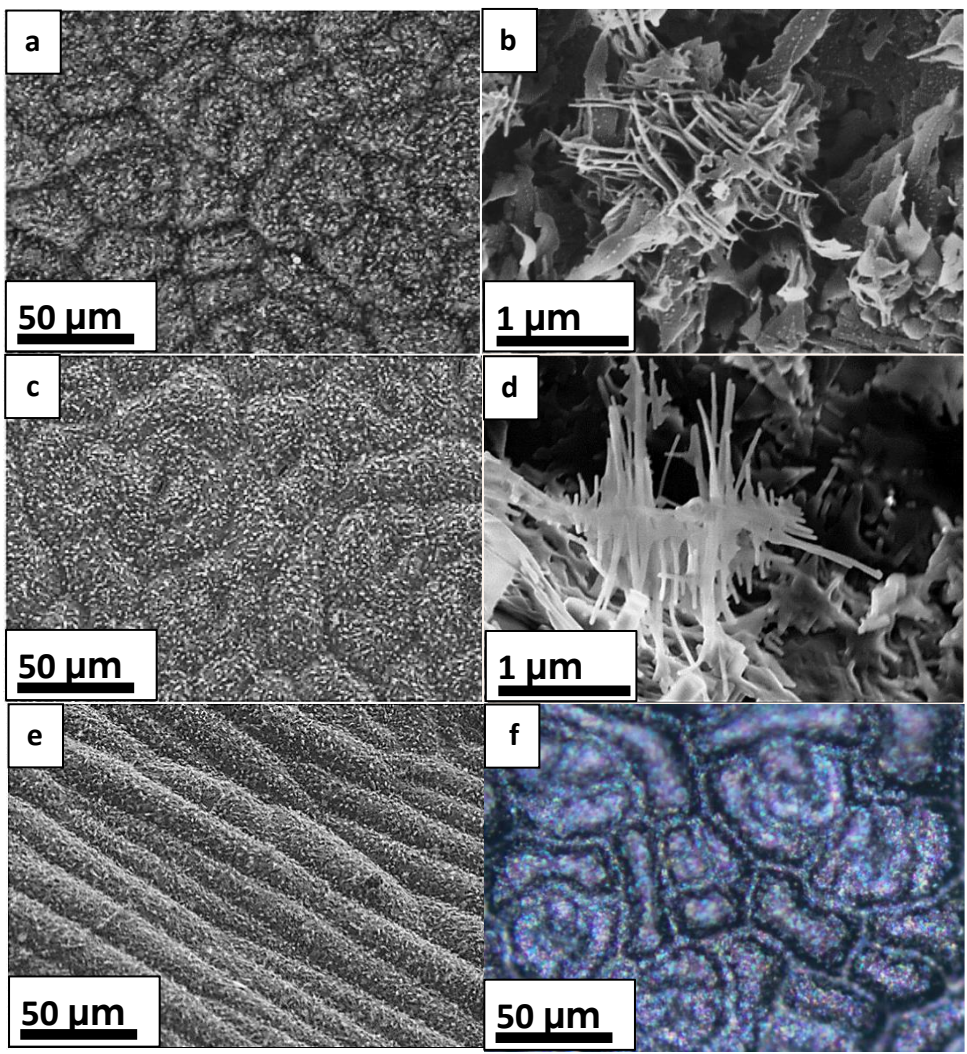

**Figure 2** - Electron microscopy and optical microscopy images of broccoli leaf surfaces. a,b) Cryo-SEM images of broccoli leaf surface (plant cells and epicuticular wax structures), c-d) HR-SEM of leaf surface after liquid substitution process, e) optical microscopy image of plant cells.

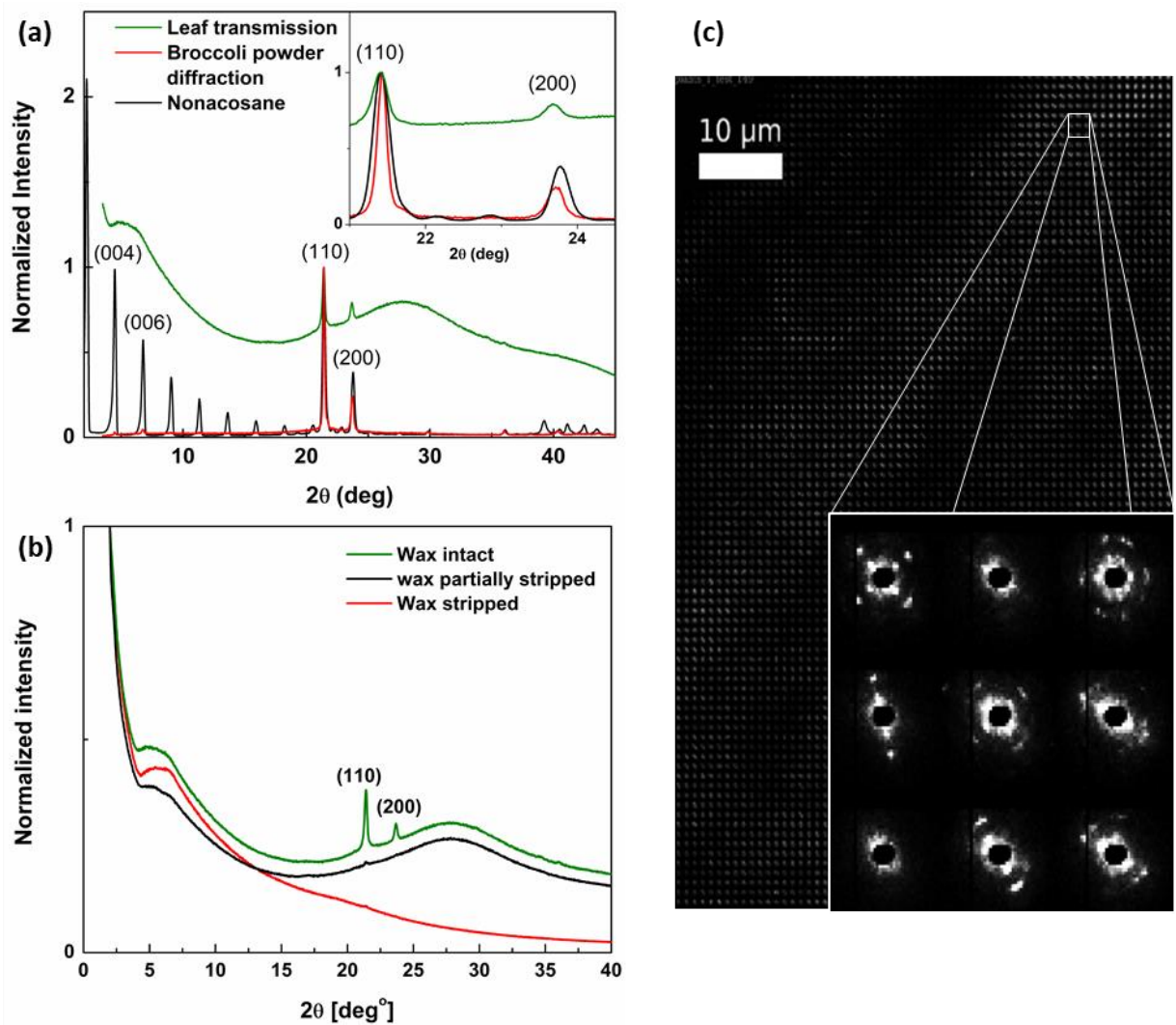

**Figure 3** – Comparison between a) undamaged leaf transmission diffraction to broccoli wax powder diffraction and pure Nonacosane, b) transmission diffractions of leaf with wax intact and wax removed (Cu kα). c) sub micron diffraction scan of the leaf. In the inset a close up of the diffraction patterns can be seen (0.3999Å).

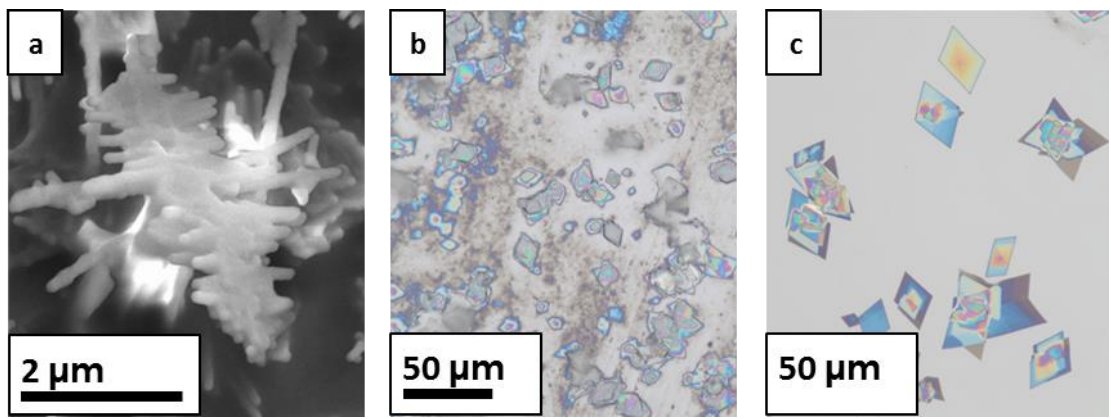

**Figure 4** - Comparison between (a) naturally occurring broccoli wax crystals –HR-SEM (b) solution deposited broccoli wax crystals – optical microscopy, and (c) solution crystalized pure nonacosane – optical microscopy.

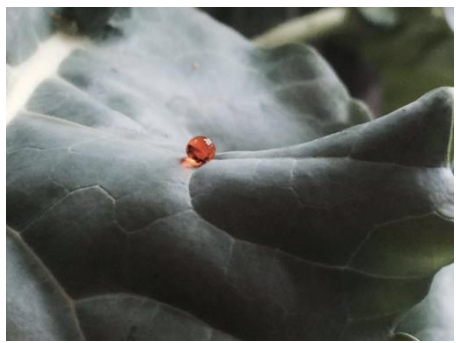

**Figure 5** - Methyl orange-stained water drop on a broccoli leaf.

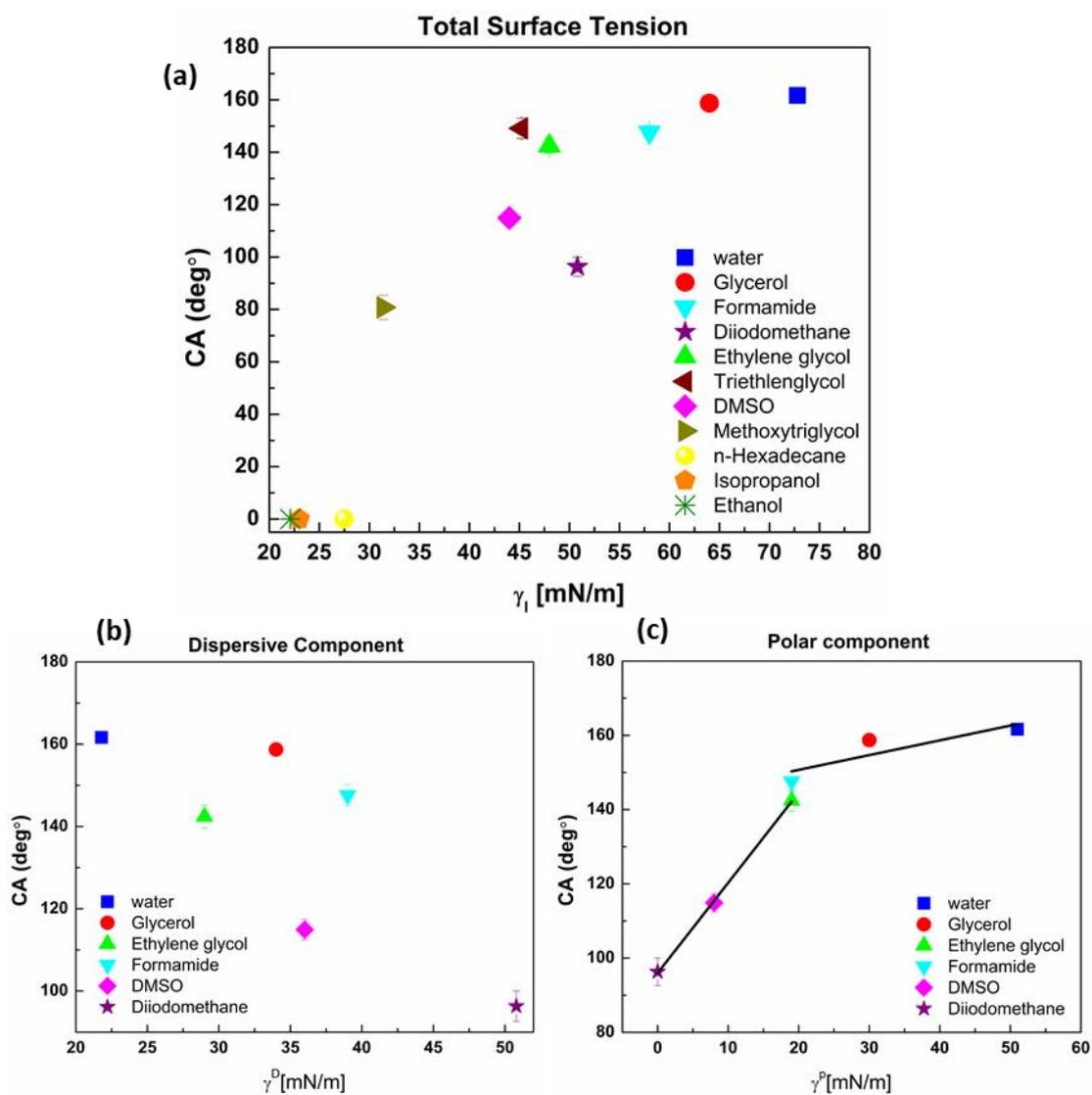

**Figure 6** - Wetting behavior versus surface tension components: a) total surface tension, b) Dispersive component, c) polar component.

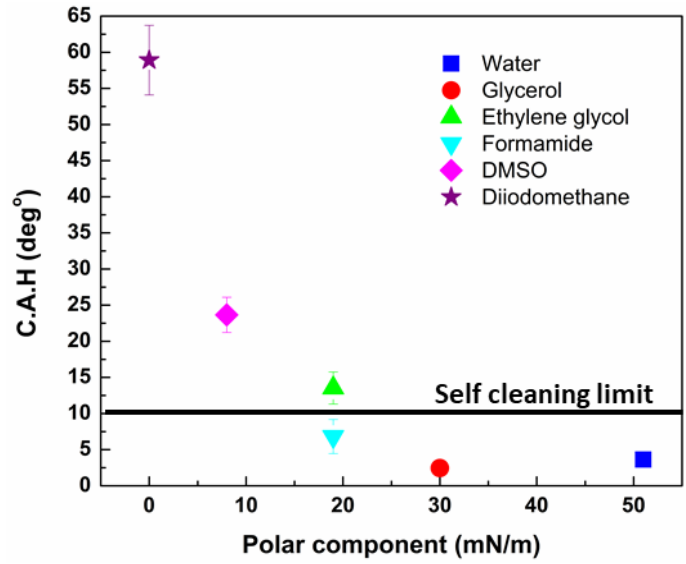

**Figure 7** - CAH versus the polar surface tension component of the different solvents.

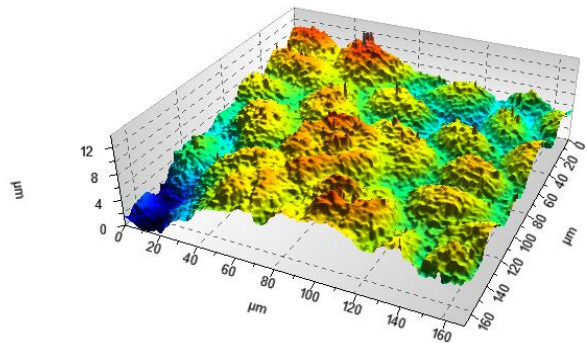

**Figure 8** – Topographical image of broccoli leaf surface.

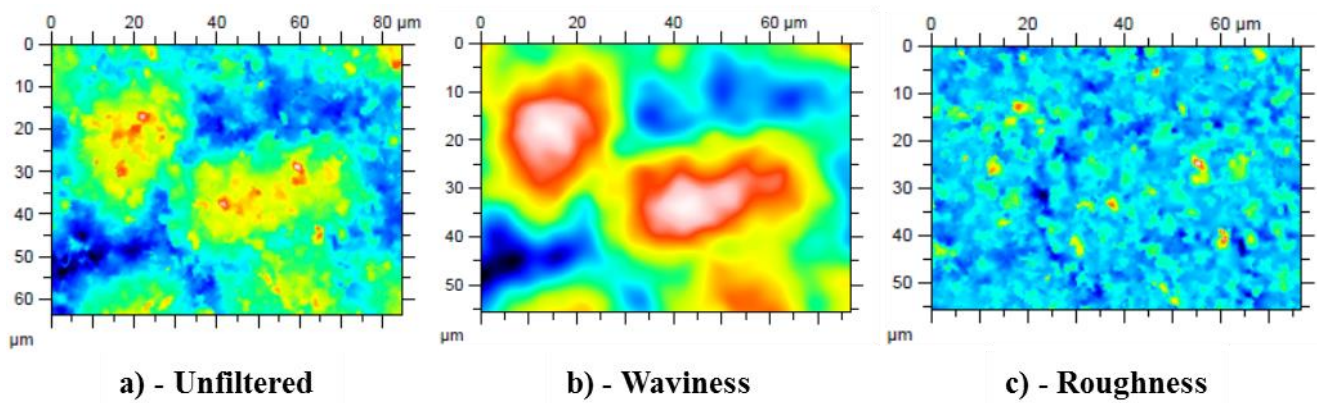

**Figure 9** - Example of filtered roughness data. a) Unfiltered data, b) macro-roughness (waviness), c) micro-roughness.

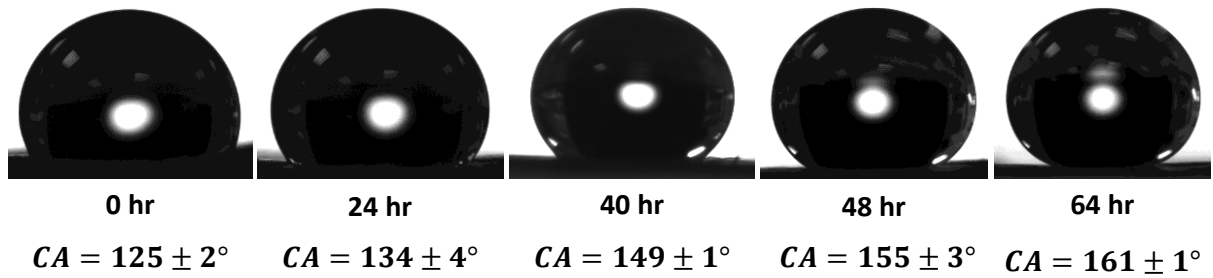

| 0 hr | 24 hr | 40 hr | 48 hr | 64 hr |
| --- | --- | --- | --- | --- |
| $CA = 125 \pm 2°$ | $CA = 134 \pm 4°$ | $CA = 149 \pm 1°$ | $CA = 155 \pm 3°$ | $CA = 161 \pm 1°$ |

**Figure 10** - Evolution of the CA of water on damaged broccoli leaf over time

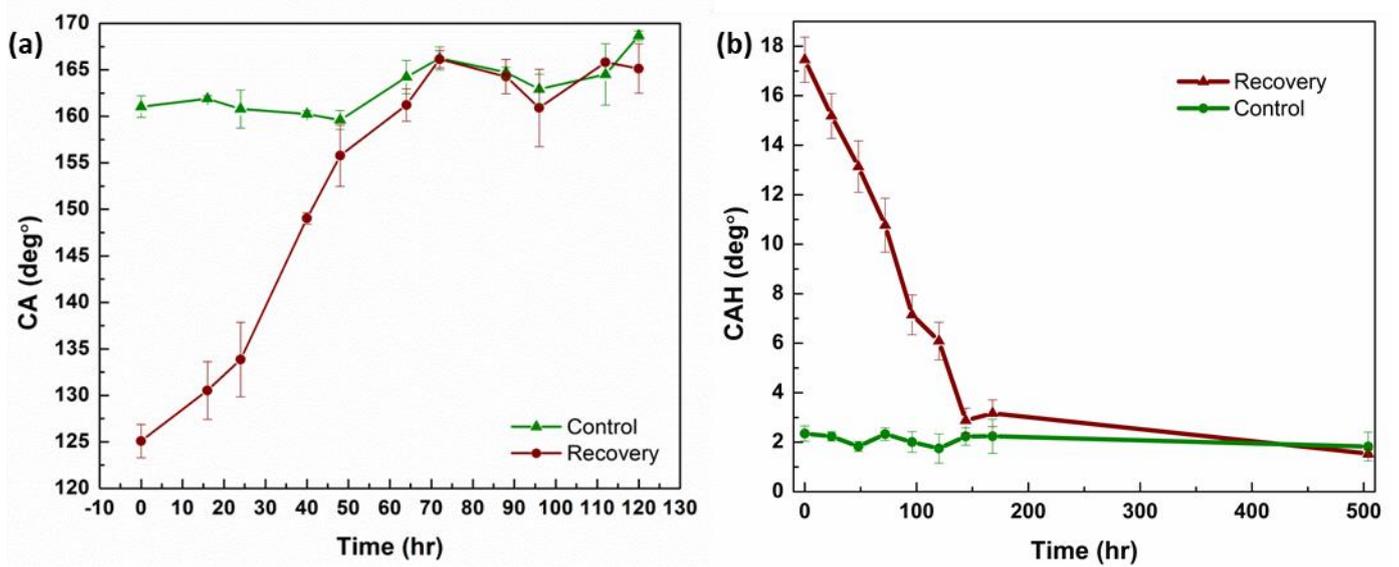

**Figure 11** - Recovery of C.A. (a) and C.A.H (b) over time.

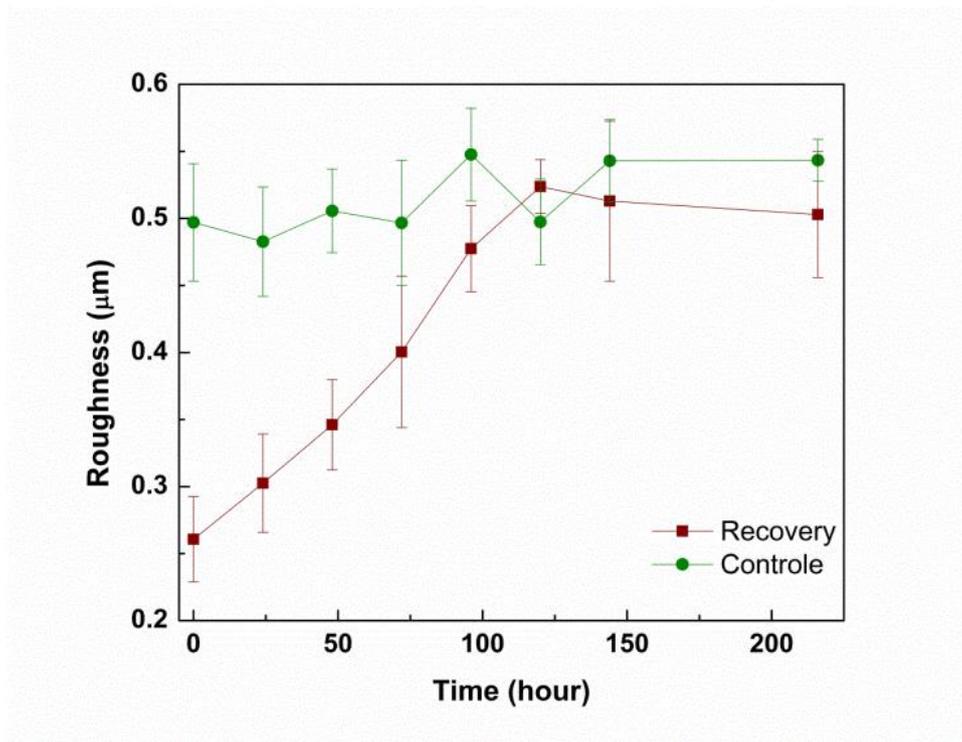

**Figure 12** - Recovery of micro-roughness after mechanical damage.

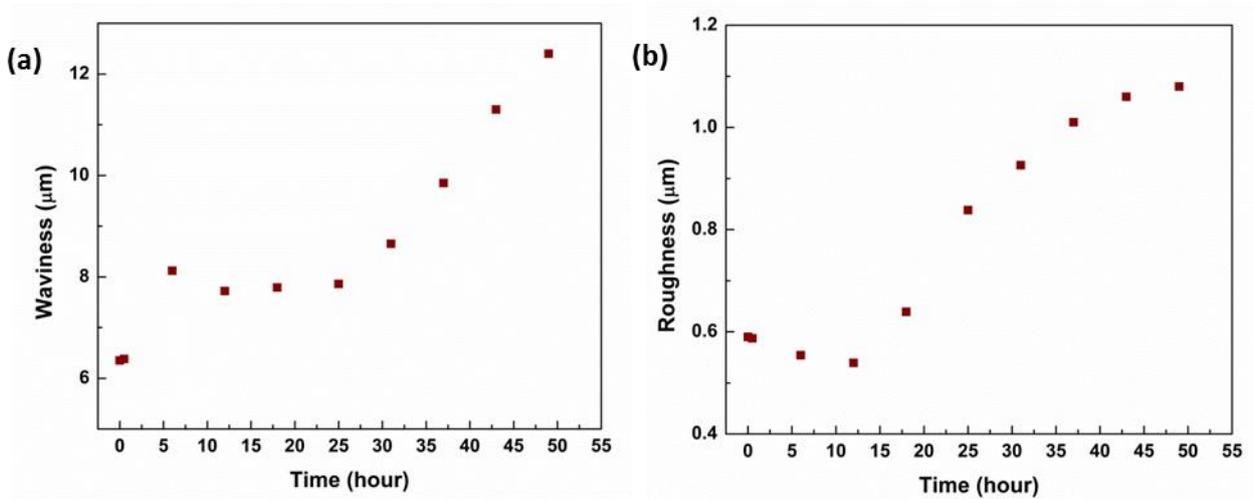

**Figure 13** - Recovery of macro-roughness via (a) waviness, (b) roughness filltering

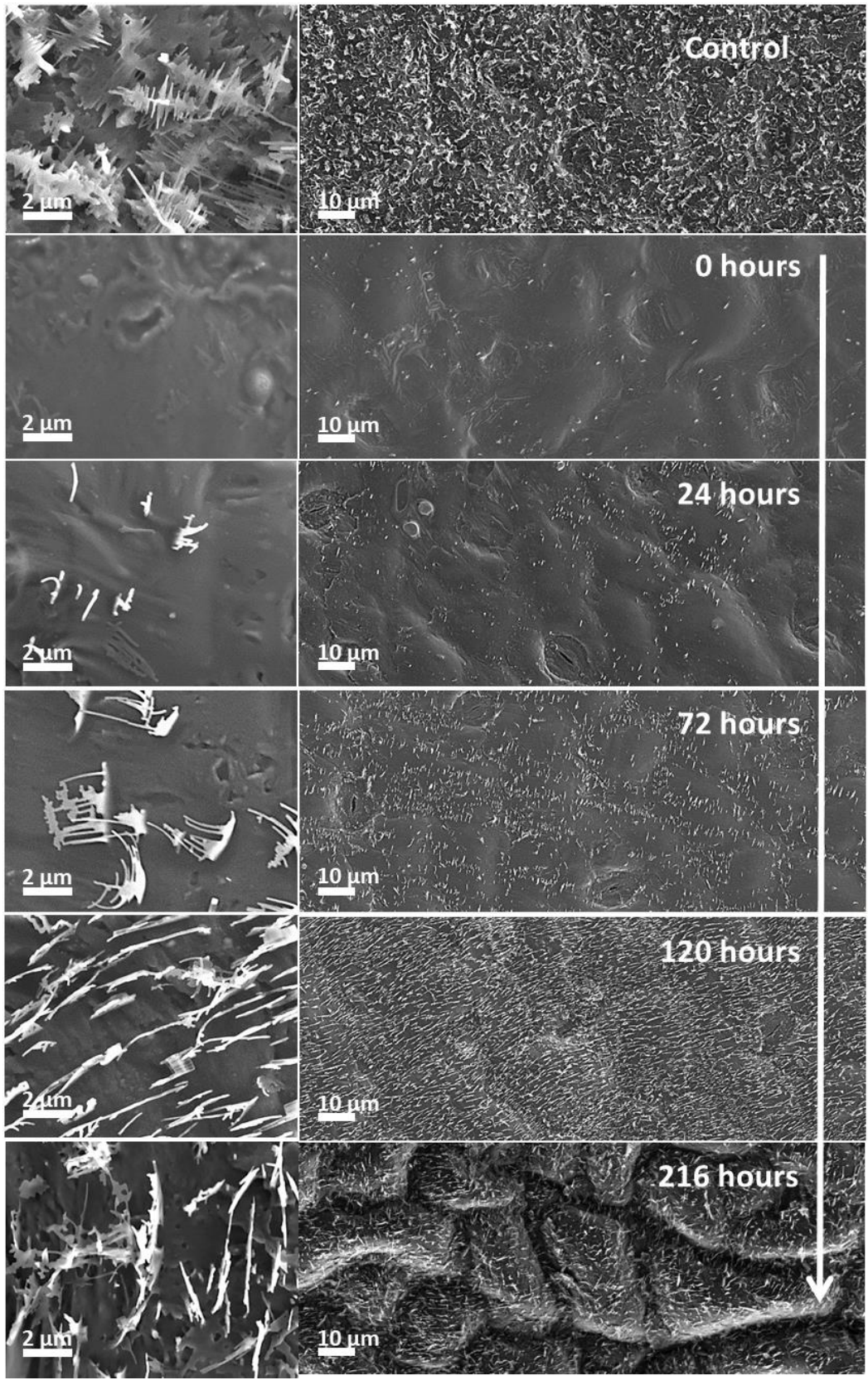

**Figure 14** - Time resolved HR-SEM of the recovery of the broccoli leaf surface

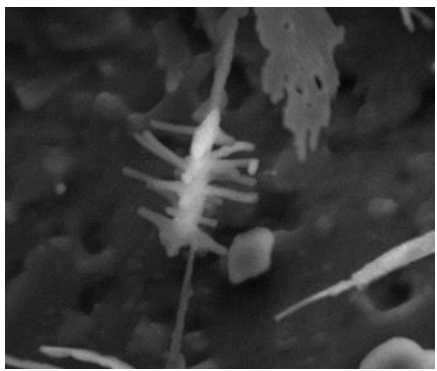

**Figure 15** - Example of complex wax structure after recovery.